\documentclass[conference]{IEEEtran}
\IEEEoverridecommandlockouts
\newcommand{\subparagraph}{}
\usepackage{titlesec}
\usepackage{amsmath,amssymb,amsfonts}
\usepackage{wrapfig}
\usepackage{graphicx}
\usepackage{setspace}
\usepackage{tabularx}
\usepackage{float}   
\usepackage{multicol}
\usepackage{indentfirst}
\usepackage{cite} 
\usepackage{caption}
\usepackage[final]{pdfpages}
\usepackage{tikz}
\usetikzlibrary{tikzmark}
\usepackage{footnote}
\usepackage{textcomp}
\usepackage{algorithm}
\usepackage{algorithmic}
\usepackage{multirow}
\usepackage{url}
\usepackage[hang,flushmargin]{footmisc}
\usepackage[super]{nth}
\usepackage{titlesec}
\usepackage{subcaption}
\usepackage{fancyhdr}
\usepackage{bm}
\usepackage{newfloat}
\usepackage{epsfig}

\usepackage{enumitem}

\DeclareFloatingEnvironment[placement={!ht},name=List]{mylist}

\def\BibTeX{{\rm B\kern-.05em{\sc i\kern-.025em b}\kern-.08em
		T\kern-.1667em\lower.7ex\hbox{E}\kern-.125emX}}

\begin{document}
	\title{Threshold Logic in a Flash
		\vspace{-12pt}}
	\author{\IEEEauthorblockN{Ankit Wagle\IEEEauthorrefmark{1}, Gian
			Singh\IEEEauthorrefmark{1}, Jinghua Yang\IEEEauthorrefmark{1},
			Sunil Khatri\IEEEauthorrefmark{2}, Sarma
			Vrudhula\IEEEauthorrefmark{1}}
		\IEEEauthorblockA{\IEEEauthorrefmark{1} (awagle1,gsingh58,jinghua.yang,vrudhula)@asu.edu, \IEEEauthorrefmark{2} sunil.khatri@tamu.edu}
		\IEEEauthorblockA{\IEEEauthorrefmark{1} School of Computing, Informatics and Decision Systems Engineering, Arizona State University, Tempe AZ 85281}
		\IEEEauthorblockA{\IEEEauthorrefmark{2} Dept. of Electrical and Computer Engineering, Texas A\&M University, College Station TX}\\[1em]
		\thanks{\IEEEauthorrefmark{1}The research was supported in part by NSF PFI award 1701241.}
		\vspace{-50pt}    
	}
	\maketitle
	%
	
	
	\begin{abstract}
		This paper describes a novel design of a threshold logic gate (a binary perceptron) and its implementation as a \textit{standard cell}.  This new cell structure, referred to as flash threshold logic (FTL), uses floating gate (flash) transistors to realize the weights associated with a threshold function.  The threshold voltages of the flash transistors serve as a proxy for the weights. An FTL cell can be equivalently viewed as a multi-input, edge-triggered flipflop which computes a threshold function on a clock edge. Consequently, it can be used in the automatic synthesis of ASICs.  The use of flash transistors in the FTL cell allows \textit{programming} of the weights after fabrication, thereby preventing discovery of its function by a foundry or by reverse engineering. This paper focuses on the design and characteristics of the FTL cell.  We present a novel method for programming the weights of an FTL cell for a specified threshold function using a modified perceptron learning algorithm.  The algorithm is further extended to select weights to maximize the robustness of the design in the presence of process variations.  The FTL circuit was designed in 40nm technology and simulations with layout-extracted parasitics included, demonstrate significant improvements in the area (79.7\%), power (61.1\%), and performance (42.5\%) when compared to the equivalent implementations of the same function in conventional static CMOS design. Weight selection targeting robustness is demonstrated using Monte Carlo simulations. The paper also shows how FTL cells can be used for fixing timing errors after fabrication.  
	\end{abstract}
	\vspace{-5pt}
	\setlist[itemize]{leftmargin=*}    
	\setlist[enumerate]{leftmargin=*}

	\section{\textbf{Introduction and Motivation}}
	\label{sec:intro}
	
	Methods to optimize the performance, power and area (PPA) of static CMOS circuits have continuously improved over three decades,  leaving few opportunities, if any, for further improvements.  This suggests that if there are to be any further advances in improving PPA at the logic and circuit levels, the conventional way of computing logic functions has to be revisited.  Although several nanotechnologies are being investigated as alternatives or enhancements to static CMOS (e.g. \cite{DSD04,berezowski:dsd05,jha_nanopipelining,Zhang_2005_ICVLSI}), they remain at the research stage and large scale adoption is still far in the future.  

	This paper introduces a new \textit{programmable ASIC primitive}, referred to as a \textit{flash threshold logic} (FTL) cell, that can be used to substantially improve all three PPA metrics of an ASIC.  An FTL cell and its use in an ASIC is different from any other type of ASIC component previously reported. However, it is designed as a \textit{standard cell}, so that it is fully compatible with conventional ASIC design flow, and can be processed by commercial design tools without any changes. In other words, it can easily be combined with conventional CMOS logic during synthesis, technology mapping, and place-and-route. However, it is functionally and structurally very different from a complex standard cell.
	
	An FTL cell of $n$ inputs can realize any \textit{threshold} function of $n$ or fewer variables. A threshold function $f(x_1, \cdots, x_n)$~\cite{Muroga} is a unate Boolean function whose on-set and off-set are \textit{linearly separable}, i.e. there exists a vector of weights $\bm{W} = (w_1, w_2, \cdots, w_n)$\footnote{W.L.O.G, weights can be assumed to be positive integers~\cite{Siu_1995}, and for a given truth table of a threshold function, there is a weight vector whose sum is minimum~\cite{Siu_1995}.} and a threshold $T$ such that \\
	\begin{equation}
	\label{eq:ThresholdDefinition}
	f(x_1, x_2, \cdots, x_n) = 1 \Leftrightarrow \sum_{i=1}^{n} w_i x_i \geq T,
	\vspace{-5pt}
	\end{equation}
	where $\sum$ here denotes the arithmetic sum.  A threshold function can be equivalently represented by $(\bm{W}, T) = (w_1, w_2, \cdots, w_n; T)$.
	
	
	
	\setlength{\intextsep}{-3pt}%
	\setlength{\columnsep}{0pt}%
	\begin{wrapfigure}[9]{r}{0.4\columnwidth}
		\centering
		\includegraphics[scale=0.48]{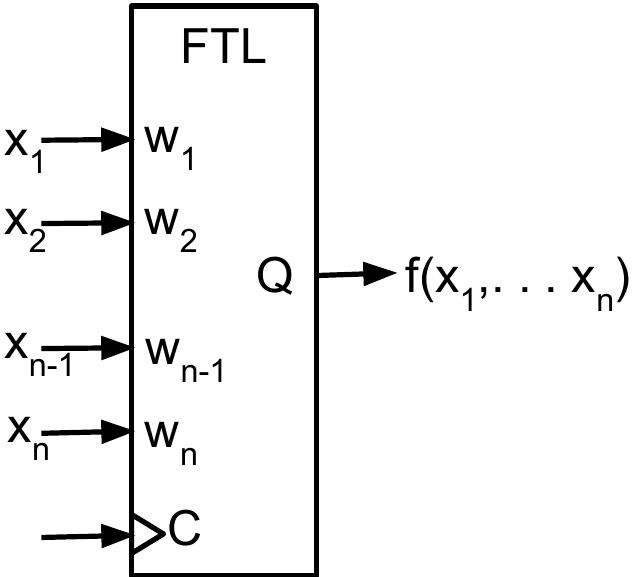}
		\caption{FTL Schematic}
		\vspace*{-30pt}
		\label{fig:FTLSchematic}
	\end{wrapfigure}
	Figure~\ref{fig:FTLSchematic} shows the schematic of the FTL cell, in which the weights $\bm{W}$ are internal parameters of the cell.  The schematic is meant to convey that the input-output behavior of an FTL cell may be viewed as an  \textit{edge-triggered}, \textit{multi-input} flip-flop, whose output is a threshold function, registered at the rising edge of the clock signal C.\\[-10pt]
	
	A distinctive characteristic of the FTL cell design is that the actual threshold function realized by an FTL instance within an ASIC is \textit{programmed after the circuit is manufactured}.  An FTL based ASIC integrates \textit{flash} or \textit{floating gate}~\cite{Cai_DATE_2013} transistors along with conventional MOSFETs within the FTL cell.  Thus, unlike many of the \textit{emerging technologies}~\cite{Perricone_DATE_2017,Yang_NANOARCH_2014,jha_nanopipelining,berezowski:dsd05}, an FTL cell employs mature IC technologies (CMOS and Flash) that can be commercially manufactured and integrated today.
	
	\subsection{\textbf{FTL in ASIC Design~--~A Valuable Use Case}}
	\label{subsec:ASICDesignUseCase}
	
	\begin{figure*}%
		\centering
		\begin{subfigure}{.3\textwidth}
			\centering
			\includegraphics[scale=0.35]{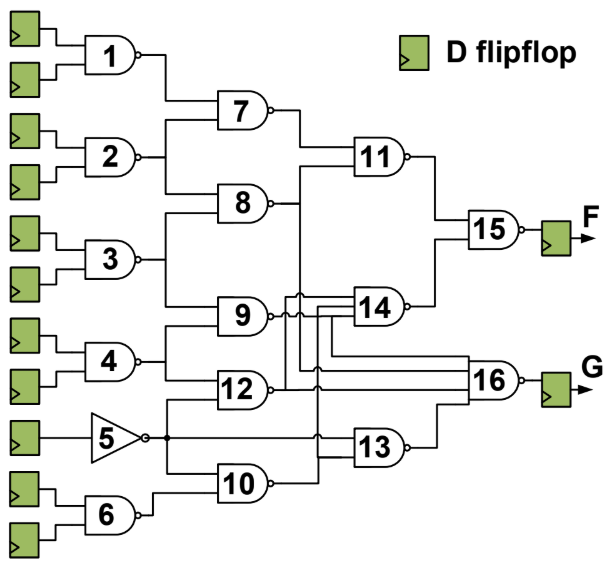}%
			\caption{A logic netlist.}%
			\label{subfiga:FTL_ASIC}%
		\end{subfigure}\hfill%
		\hspace{-30pt}
		\begin{subfigure}{.30\textwidth}
			\centering
			\includegraphics[scale=0.35]{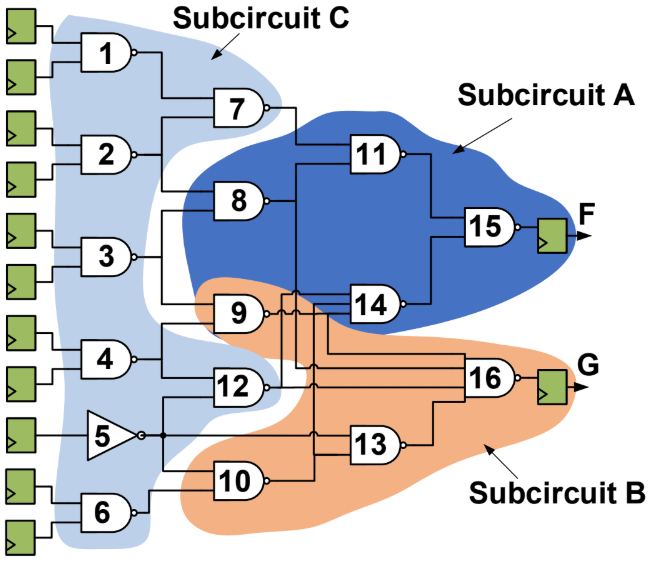}%
			\vspace{-5pt}
			\caption{Identifying threshold functions in TFI cones of flip-flops}%
			\label{subfigb:FTL_ASIC}%
		\end{subfigure}\hfill%
		\begin{subfigure}{.3\textwidth}
			\centering
			\includegraphics[scale=0.37]{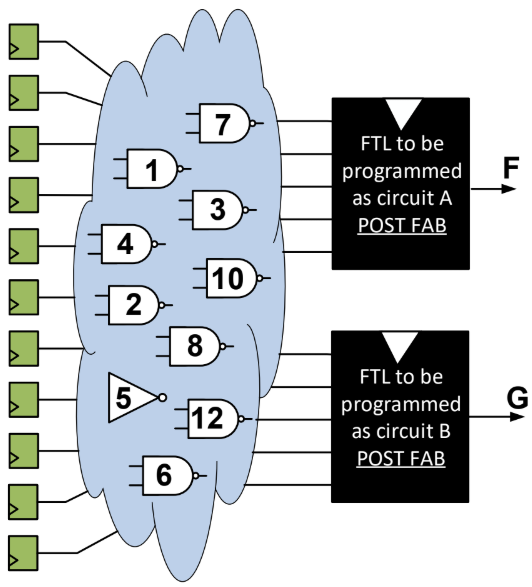}%
			\caption{A FTL-CMOS logic hybrid}%
			\label{subfigc:FTL_ASIC}%
		\end{subfigure}%
		\vspace{-10pt}
		\caption{Use of FTL in ASIC design}
		\label{fig:FTL_ASIC}
		\vspace*{-20pt}
	\end{figure*}
	
	The focus of this paper is on the design of the FTL cell.  Before proceeding to that, it will be instructive to understand its use in ASIC design\cite{Kulkarni_TVLSI_2016}. The fact that an FTL is a programmable, multi-input flip-flop provides a unique and significant new opportunity to improve the PPA of ASICs.   
	
	Consider the logic netlist shown in Figure~\ref{subfiga:FTL_ASIC} which has two registered outputs $F$ and $G$. Suppose that transitive fan in (TFI) cones of $F$ and $G$ are traversed and two subcircuits $A$ and $B$ (see Figure~\ref{subfigb:FTL_ASIC}) are found that are threshold functions of their inputs. The remaining subcircuit is labeled as $C$.  Suppose that subcircuits $A$ and $B$ are each replaced by an FTL cell, programmed to realize $A$ and $B$.  This replacement is shown in Figure~\ref{subfigc:FTL_ASIC}, where the FTL cells are shown as black boxes.  Now, subcircuit $C$ would be re-synthesized to account for the changes in the delay of FTL cells and the new loads that they present to the outputs of $C$. The circuit in Figure~\ref{subfigc:FTL_ASIC} would substantially improve the PPA of an ASIC for two reasons:
	\begin{enumerate}
		\item Subcircuits $A$ and $B$ and the two flip-flops are each replaced by an FTL cell which has much few transistors, resulting in a significant reduction in area and power. 
		\item The clock-to-Q delay of FTL cells are typically about 30\% to 40\% smaller than the delay of standard cell realization of subcircuits $A$ and $B$ plus the clock-to-Q delay of regular flip-flops. In the FTL-CMOS hybrid design, this results in a substantial amount of slack (required time minus arrival time) on the outputs of subcircuit $C$, which in turn will allow synthesis and technology mapping tools to drastically reduce the logic area of subcircuit $C$.  
	\end{enumerate}
	
	FTL based ASICs also offers several other equally significant advantages not possible with conventional CMOS logic.  
	\begin{enumerate}
		\item \textit{IP Protection:} A CMOS ASIC with embedded FTL cells cannot be reverse-engineered by a foundry or any third party because the functions of the FTL cells are unknown (black boxes) at manufacturing time, as shown in Figure~\ref{subfigc:FTL_ASIC}.
		
		\item \textit{Correcting Timing Errors:} The fine-grained, post-manufacture flash threshold voltage programmability allows precise speed binning, and correction of timing errors.  This is not possible in traditional CMOS design.
		
		\item \textit{Mitigating Aging Effects:} By re-programming the flash design in-field, our scheme allows for mitigating the effects of aging. This is also not possible in CMOS design.
		
		\item \textit{High Endurance:} Unlike flash memory, the FTL cell does not suffer from endurance issues. Flash transistors can endure a finite number of write cycles (1K to 100K) \cite{Jung_CASES_2007}, \cite{Boboila_2010}. In our approach, the flash devices will be programmed a few times (at most), after fabrication, and then again to possibly adjust for aging effects (in the field).
	\end{enumerate}
	
	\subsection{\textbf{Main Contributions}}
	
	The remainder of the paper will focus on the design of the FTL cell and demonstrate its key characteristics through extensive and detailed electrical simulations using the state-of-the-art device and circuit models and commercial tools.  The main contributions of this work are summarized below. 
	\begin{itemize}[noitemsep,topsep=0pt]
		\item This paper introduces a  novel circuit design of the FTL cell to realize all threshold functions of $n$ or fewer variables\footnote{In the experimental results, we find that $n$ = 5 is a sufficiently good choice to demonstrate substantial improvements in PPA, since there are a large number (117) of threshold functions of $n$ or fewer variables}.  The new design incorporates both flash transistors and conventional MOSFETs in a unique way to realize highly robust threshold logic circuits.  
		
		\item The set of threshold voltages ($\bm{V}_t$) of the flash transistors in the FTL cell serve as a proxy for $[\bm{W},T]$ that define a threshold function realized by an FTL cell. Since the threshold voltages of the flash transistors can be programmed with high precision~\cite{Cai_DATE_2013}, an FTL cell can implement weights with great fidelity.  We introduce an algorithm that maps the weights of a given threshold function $f = [\bm{W},T]$ to the threshold voltages of the flash transistors.  This is a complex, non-linear, multi-valued mapping. That is, several different $\bm{V_t}$(s) may correspond to a given $\bm{W,T}$, each determined by the complex electrical and layout characteristics of the MOSFETs and flash transistors.  Given a layout extracted netlist of an FTL cell, we present a novel modification of the classical \textit{perceptron learning algorithm} (PLA)~\cite{rosenblatt_PR_1958} that works in concert with HSPICE to determine one $\bm{V_t}$ of an FTL cell that computes $f = [\bm{W},T]$. This algorithm accounts for layout parasitics and process variations. Like the original PLA, the modified PLA is guaranteed to converge, ensuring that a solution ($\bm{V_t}$) for the given layout of an FTL cell will be found in a finite number of steps if a solution exists. 
		
		\item The fine-grained programmability of threshold voltages of the flash transistors in an FTL cell is exploited to improve its robustness.  Given that the mapping $[\bm{W}, T] \Rightarrow \bm{V}_t$ is multi-valued, we show how to direct our modified PLA to find a $\bm{V}_t$ that will ensure that the FTL cell reliably computes the given threshold function in the presence of local and global process and environmental variations. Using this approach, substantial improvement in the robustness of the FTL cell is demonstrated using Monte Carlo simulations.  This also shows how post-fabrication tuning of the threshold voltages can correct failures due to process variations, or modify the delay to correct timing errors, improve a circuit's performance, or improve the performance characteristics of a design to alter the speed binning distribution in a manner that maximizes profit.
	\end{itemize}
	
	\subsection{\textbf{Organization of the Paper}} 
	
	Section \ref{sec:background} gives a very brief overview of threshold logic and flash transistor technology. Sections \ref{sec:FTLArchitecture}, \ref{sec:PLA} and \ref{sec:ExpResults} contain the main body of this work.  The architecture and operation of the FTL cell are described in Section~\ref{sec:FTLArchitecture}.  This is followed by a description in Section~\ref{sec:PLA} of the modified PLA used to program an FTL to implement a given threshold function.  Section~\ref{sec:ExpResults} contains an extensive set of experimental results, demonstrating the significant improvements in PPA of FTL cells over their CMOS equivalents, and validating several of the uses of post-fabrication programming/tuning of the flash devices.  Before concluding the paper in Section~\ref{sec:Conclusions}, we present a brief and partial review of the prior art related to this paper in Section~\ref{sec:RelatedWork}. 
	
	\section{\textbf{Background}}
	\label{sec:background}
	
	\subsection{\textbf{Threshold Logic}}
	
	Equation~(\ref{eq:ThresholdDefinition}) defines an $n$-input threshold function.  An example of a 5-input threshold function is a 3-out-of-5 majority function: $f(a,b,c,d,e) = abc \vee abd \vee abe \vee acd \vee ace \vee ade \vee bcd \vee bce \vee bde \vee cde \equiv a+b+c+d+e \geq 3 \equiv  [w_a, w_b, w_c, w_d, w_e; T] = [1, 1, 1, 1, 1; 3]$.  An XOR is a simple example of a non-threshold function. The importance of threshold logic stems from the fact that many Boolean functions that require exponential size AND/OR networks can be realized by polynomial sized, fixed depth threshold networks\cite{Siu_1995}. From a practical perspective, nearly 70\% of the functions in standard cell libraries are threshold functions. We will demonstrate that implementing threshold functions using conventional CMOS logic primitives is very inefficient, as compared to the FTL cell. In our approach, Equation~(\ref{eq:ThresholdDefinition}) must be translated to the comparison of electrical quantity such as charge, current or voltage. This is the basis of many other threshold gate implementations as well~\cite{beiu:2003}.
	
	\subsection{\textbf{Flash Transistors}}
	
	\setlength{\intextsep}{-8pt}%
	\setlength{\columnsep}{6pt}%
	\begin{wrapfigure}[11]{r}{.5\columnwidth}
		\centering \includegraphics[scale=.18]{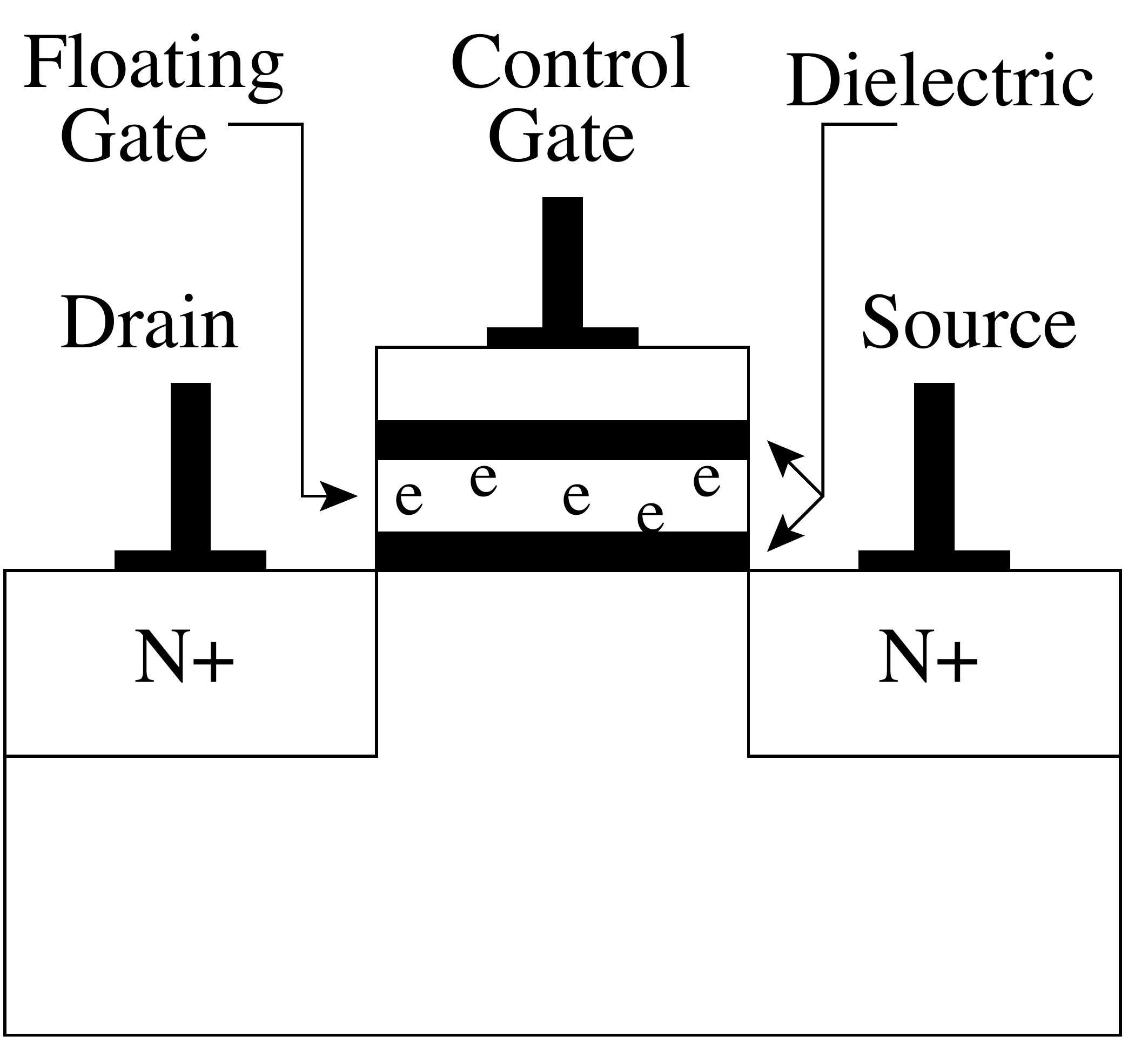}
		\caption{Flash Transistor Cross Section}
		\label{fig:flash}
	\end{wrapfigure}
	Flash or floating gate transistors are dual-gate field effect transistors (DGFETs). The first gate is called a {\em control gate} and the second is a {\em floating gate} (see Figure~\ref{fig:flash}). The control gate is similar to the gate of a traditional MOSFET. The floating gate is inserted between the substrate and the control gate, and is electrically and physically isolated.  Hence, current cannot flow into (out) of the floating gate, unless electrons are forced to enter (leave) the floating gate from (to) the substrate by a phenomenon known as Fowler-Nordheim (FN) tunneling~\cite{Fowler_RSL_1928}. 
	
	A flash device is programmed by holding its body, source and drain nodes at the ground and applying a high voltage (10-20 Volts) to the control gate. The resulting electric field forces electrons to tunnel from the substrate into the floating gate, increasing the threshold voltage of the flash transistor. The resulting threshold voltage depends on the number of electrons that tunnel into the floating gate, which depends on the duration of the programming pulse. Significantly, the threshold voltage of a flash transistor can be adjusted with a fine granularity~\cite{Cai_DATE_2013}. Once electrons are trapped in the floating gate, they remain trapped for many years~\cite{Jung_CASES_2007, Boboila_2010}, or until removed by an erase operation. A flash transistor can be erased by holding the control gate to ground, floating the drain and source nodes, and applying a high voltage at the body node. Erasing is simultaneously performed on all the transistors which share a common body node.

	\begin{figure*}[!htbp]
		\begin{minipage}[c]{0.66\textwidth}
			\includegraphics[width=\columnwidth]{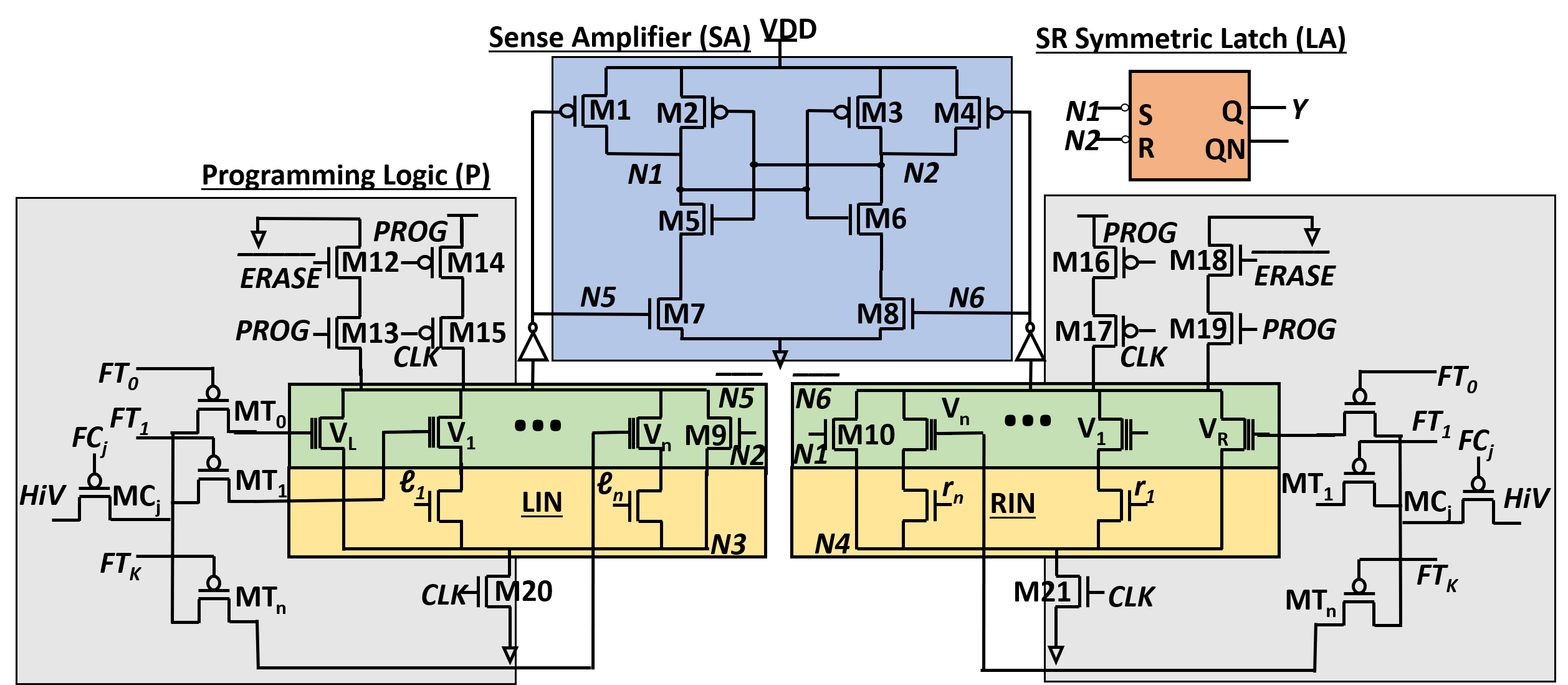}
		\end{minipage}\hfill
		\begin{minipage}[c]{0.3\textwidth}
			\hspace{-30pt}
			\vspace{10pt}
			\caption{\small FTL Cell Architecture: Input networks LIN and RIN drive the sense amplifier with current based on weighted inputs. Input weights are implemented by modulating the conductivity of LIN and RIN using flash transistors. Sense amplifier  evaluates the threshold function and drives the latch to produce the output Y. An FTL cell is programmed by sending high voltage pulses to the flash transistors' gates via  the Programming Logic.}
			\label{fig:flash_tlg}
		\end{minipage}
		\vspace{-15pt}
	\end{figure*}

	
	\section{\textbf{Flash Threshold Logic (FTL) Cell}}
	\label{sec:FTLArchitecture}
	Figure~\ref{fig:flash_tlg} shows the architecture of the FTL cell.  It has five main components: the left input network LIN, the right input network RIN, a sense amplifier (SA), an output latch (LA) and a flash transistor programming logic (P).  The LIN and RIN consist of two sets of inputs $(\ell_1, \cdots, \ell_n)$ and $(r_1, \cdots, r_n)$, respectively, with each input in series with a flash transistor. In our implementation, $\ell_i=\overline{r_i}$ for all $i$. The conductivity of these two networks is determined by the state of the inputs and the threshold voltages of the flash transistors. The assignment of signals to the LIN and RIN is done to ensure sufficient difference in conductivity across all minterm pairs ($m_i,m_j$) such that $f(m_i)\ne f(m_j)$. 
	
	The FTL cell has two differential signals $N1$ and $N2$, which serve as inputs to an SR latch.  When $[N1, N2] = [0,1]$ ($[1,0]$), the latch is set (reset) and the output $Y = 1 (0)$.  The magnitudes of the two sides of the inequality~(\ref{eq:ThresholdDefinition}) in the definition of a threshold function are mapped to the \textit{conductance} $G_L$ of the LIN and $G_R$ of the RIN, such that $[N1, N2] = [0,1] \Leftrightarrow G_L > G_R$ and $[N1, N2] = [1,0] \Leftrightarrow G_L < G_R$.  As stated earlier, the flash transistor threshold voltages serve as a proxy to the weights of the threshold function~--~the higher the weight, the lower will be the threshold voltage. For a given threshold function, this non-linear monotonic relationship is \textbf{\textit{learnt}} using a modified perceptron learning algorithm described in Section~\ref{sec:PLA}. 
	
	The FTL cell has three modes: \textit{regular}, \textit{erase} and \textit{programming} mode. The $\bm{V}_t$ values of the flash transistors are set in the programming mode and erased in the erase mode. The evaluation takes place in regular mode. 
	
	\noindent \textbf{FTL Regular Mode:} In this mode $\text{PROG} = \text{ERASE} = 0$. Assume that the $\bm{V}_t$s of the flash transistors have been set to appropriate values corresponding to the weights of the threshold function, and their gates are being driven to 1 by setting HiV to VDD, $FC_j$ to 0V and all $FT_i$ to 0V. When $CLK = 0$, the circuit is reset.  In this phase, the nodes $\overline{N5}$ and $\overline{N6}$ of LIN and RIN are connected to the supply, $N5 = N6 = 0$, and $N1 = N2 = 1$. Therefore, the output $Y$ remains unchanged. 
	
	Assume now that an on-set minterm is applied to the inputs in the LIN and RIN.  With properly assigned $\bm{V}_t$ values to the flash transistors, suppose that $G_L > G_R$ for the given minterm.  When  $CLK: 0 \rightarrow 1$, both the LIN and RIN will conduct, and $N5$ and $N6$ will both transition from $0 \rightarrow 1$. Assuming $G_L > G_R$, $N5$ rises faster than $N6$, and hence $N5$ will make $M7$ active before $N6$ makes $M8$ active. This will start to discharge $N1$ before $N2$. When $N1$ falls below the $V_t$ of $M6$, it will stop further discharge of $N2$, and turn on $M3$, resulting in $N2: 0 \rightarrow 1$.  Finally, [N1,N2] = [0,1] sets the SR latch, resulting in $Y = 1$.  For an off-set minterm, $G_L < G_R$, and $[N1, N2]  = [1,0]$ resulting in $Y = 0$. 
	
	The conventional circuit structures used in flash memories are not suitable for programming an FTL cell because it has to also perform logic operations.  Consequently, we present a new programming interface for an off-chip programming circuit to set the $V_t$ values of any FTL cell. During flash-programming, this interface uses the $FC_j$ signal to select the $j^{th}$ FTL cell and the $FT_i$ signal to select the $i^{th}$ flash transistor of the selected FTL cell. 
	
	
	
	\noindent \textbf{FTL Programming Mode}:(ERASE=0, PROG=1, CLK=0, FT$_i$=0, FC$_j$=0, HiV=20V). The ERASE and PROG signals turn on M12 and M13 and turn off M14. In this state, the source of the flash transistor is floating while the drain and bulk are connected to the ground. Activating the appropriate transistors using the $FT_i$ and $FC_j$ signals, high voltage pulses are passed on the HiV line through $MC_j$ and $MT_i$ to the gate of the flash transistor to set the desired threshold voltage ($V_t$).
	
	\noindent \textbf{FTL Erase Mode}: (ERASE=1, PROG=1, CLK=0, FT$_i$=0, FC$_j$=0, HiV=-20V). M12 is turned off by the ERASE signal. Both the source and drain of the flash transistors are floating in this state, while the bulk is connected to the ground. A negative HiV pulse at the gate terminal of all the flash transistors in this state will tunnel the charge from the floating gate, thereby erasing the flash transistor. 

	\section{\textbf{Modified Perceptron Learning Algorithm}}
	\label{sec:PLA}
	
\begin{figure}[]
		\centerline{\includegraphics[width=0.8\columnwidth]{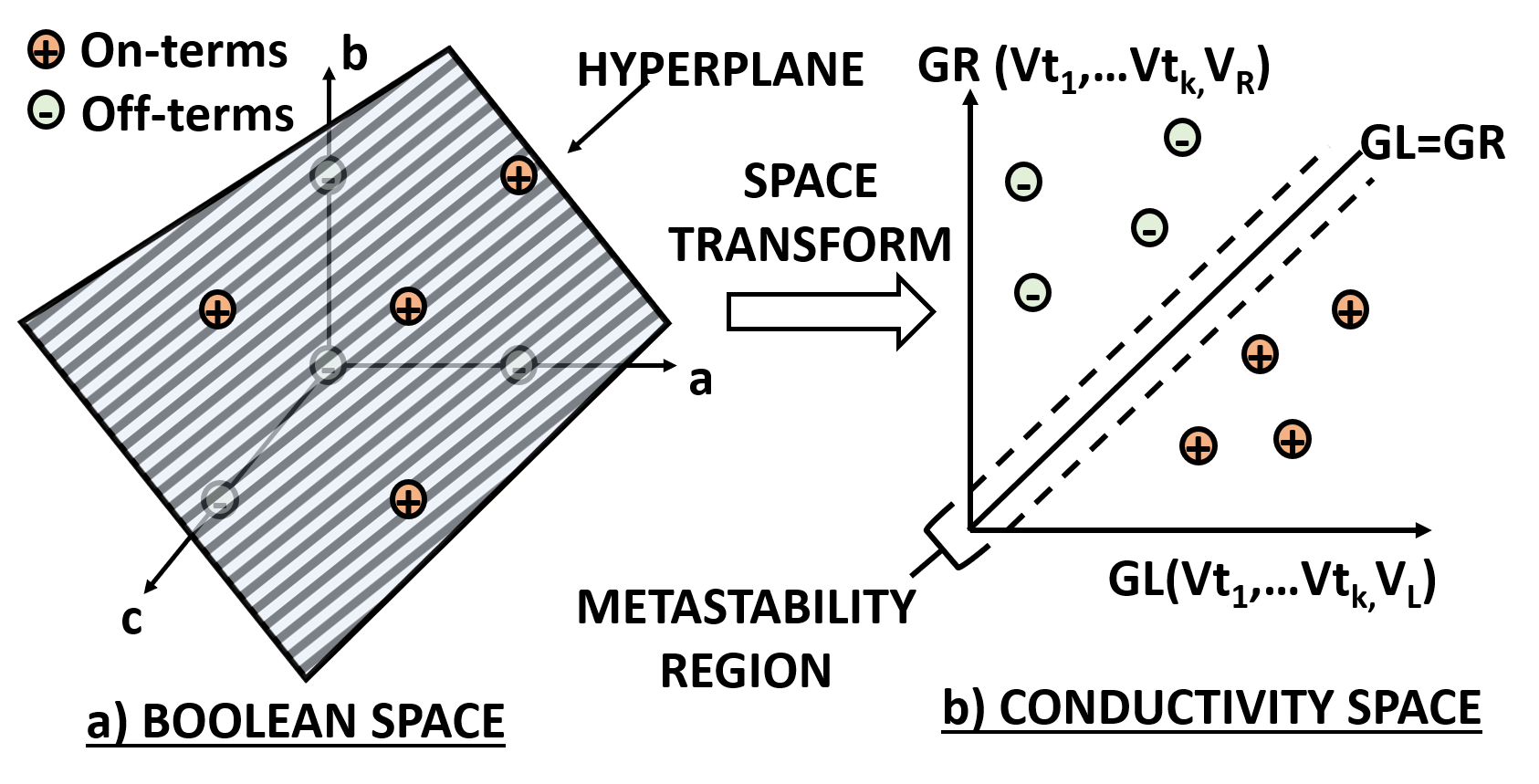}}
		\caption{\small Transformation from Boolean space to conductivity space; Hyperplane gets converted into a line.}
		\label{fig:space_transformation}
		\vspace{-21pt}
	\end{figure}

	In this section, we describe an algorithm to determine the vector of flash transistor threshold voltages for a given threshold function $f = [\bm{W},T]$. The problem is to find a mapping between the Boolean space $B^n$, and the conductivity space $(G_L, G_R)$ such that $G_L > G_R$ iff $\sum w_i x_i > T$ (i.e. for an on-set minterm), and $G_L < G_R$ iff $\sum w_i x_i < T$ (i.e. for an off-set minterm). This mapping is depicted in Figure~\ref{fig:space_transformation}. $G_L$ and $G_R$ are non-linear functions of the flash transistor threshold voltages, the time-varying drain and sources voltages of the input transistors, and the layout parasitics that vary from instance to instance.  To account for these dependencies, $G_L$ and $G_R$, in principle, must be obtained by solving a set of differential equations~--~an approach that is not practical. We next show how to simultaneously solve the differential equations numerically and perform the binary classification by a modified version of the classical perceptron learning algorithm (PLA)~\cite{rosenblatt_PR_1958}.  
	
	The PLA starts with an initial hyperplane in the Boolean space and iteratively adjusts it until all the on-set and off-set minterms fall on opposite sides of the hyperplane. Each minterm corresponds to some point in the $(G_L, G_R)$ space. Our modified PLA iteratively adjusts the $V_t(s)$ of flash transistors such that points in the conductivity space that correspond to the on-set and off-set minterms fall on the appropriate side of the line $G_L = G_R$ (Fig. \ref{fig:space_transformation}). We use HSPICE to determine whether any point falls above or below this line.
	
	A description of the modified PLA follows. The threshold voltages of the flash transistors associated with the input transistors in the LIN and RIN are labeled $V_1, V_2, \cdots, V_n$. The $i^{th}$ transistor in both LIN and RIN has a threshold voltage $V_i$. In addition, there are two special flash transistors, whose threshold voltages are $V_L$ and $V_R$ associated with the LIN and RIN, respectively.  For a threshold function $f = (w_1, w_2, \cdots, w_n; T)$, the $V_i$, $1 \leq i \leq n$, correspond to the weights $w_i$ of a threshold function, whereas only one of $V_L$ or $V_R$ is associated with the threshold $T$ of $f$.  If $V_L$ is associated with $T$, then $V_R = V_{DD}$, effectively turning it off. If $V_R$ is associated with $T$, then $V_L = V_{DD}$. The use of additional flash devices on both sides of the FTL cell allows for extra programming flexibility. The induced symmetry also balances the parasitics of the LIN and the RIN.
	
	
	For the truth table ($TT$) of $f$, the modified PLA applies all the minterms of $f$ to the FTL cell, and records the HSPICE response in an array called $OT$ (output table). For a given minterm $m_i$, if $TT(m_i) = OT(m_i)$ then the response is called a \textit{correct} response, otherwise it is called an \textit{incorrect} response. An FTL cell is completely programmed if the recorded response for every minterm is \textit{correct}. Until the FTL cell is completely programmed, at least one minterm would generate an \textit{incorrect} response. In the event of an \textit{incorrect} response associated with minterm $m_i$, the modified PLA adjusts the threshold voltages of all flash transistors associated with the ON input transistors within the interval $[\delta, V_{DD} - \delta]$, by a minimum increment $\delta$, using the following equations (k denotes the iteration number of the algorithm):
	\begin{eqnarray}
	\label{Eq:vt_update}
	\small
	V_{i}^{k+1} = \left\{ \begin{array}{ll}
	V_{i}^{k} - \delta m_i & m_i \cdot W \ge T \\
	V_{i}^{k} + \delta m_i & m_i \cdot W < T.
	\end{array}
	\right.
	\end{eqnarray}
	
	Equation~(\ref{Eq:vt_update}) is quite easy to understand. The term $\delta m_i$ is simply a vector which has a value $\delta$ at all locations where $m_i$ is 1, and zero elsewhere. For instance, $\delta (1, 0, 1, 1, 0) = (\delta, 0, \delta, \delta, 0)$.  Suppose $m_i$ is an on-set minterm for which the response was incorrect. This means that $G_L < G_R$.  Therefore $G_L$ needs to be increased for minterm $m_i$. Hence the threshold voltages of all flash transistors that are connected to the input transistors that are ON for minterm $m_i$,  should be decreased by $\delta$. Similarly, if $m_i$ is an off-set minterm, then the threshold voltages of the same flash transistors must be increased by $\delta$. This is what is expressed in Equation~(\ref{Eq:vt_update}). 
	
	Since the $V_i$ values are bounded above and below, it might not be possible to satisfy the truth table using the $V_{i}$ alone. In such cases, the algorithm will resort to adjusting $V_L$ and $V_R$ using the same principle as in Equation~(\ref{Eq:vt_update}). If $m_i$ is a on-set minterm that was incorrect, then $G_R$ should be reduced.  Therefore, $V_R$ is incremented by $\delta$, until its upper bound is reached. If this is not sufficient, then $G_L$ has to be increased.  Hence,  $V_L$ is decremented.  Given a threshold function and a sufficiently small $\delta$, the modified PLA will converge to a feasible threshold voltage set assignment $\bm{V}_t^*$ for the FTL cell~\cite{rosenblatt_PR_1958}. For an $n$-input threshold function, a pessimistic upper bound on the number of iterations is given by $kmax = 2n||\bm{V}_t^{*}||^2/\delta^2$.  For $n = 5$ and $\delta = .02V$, $kmax = 2500||\bm{V}_t^{*}||^2$. 
	
	\subsection{\textbf{Training for Robustness}}
	\label{robustness_training}
	
	The modified PLA does not consider the relative location of the points with respect to the metastability region around the line $G_L = G_R$ (see Figure~\ref{fig:space_transformation}b). Even though minterms are classified correctly, they can be arbitrarily close to the line.  The further away a minterm is from the line, the easier (and faster and more robust) it will be for the sense amplifier to detect the difference between $N5$ and $N6$, and discharge the appropriate side ($N1$ or $N2$) first.  Our approach to making the FTL cell highly robust is to introduce an additional capacitance $C_1$ on node $N1$ when classifying an on-set minterm,  and determining the maximum value of $C_1$ for which the modified PLA converges.  This \textit{handicaps} node $N1$ and directs the algorithm to find a solution, which will result in increasing $G_L$ more than increasing $G_R$.  Similarly, we add a capacitance $C_0$ on node $N2$, when classifying an off-set minterm. The corresponding threshold voltages found by the modified PLA algorithm will increase the gap between $G_L$ and $G_R$, which makes it much more robust, and also improves its speed, as a direct consequence. Note that $C_0$ and $C_1$ are introduced in the simulations for improving the training solution only, and are not part of the FTL cell.

	\section{\textbf{Experimental Results}}
	\label{sec:ExpResults}
	
	\subsection{\textbf{Experiment Setup}}
	
	A 5-input FTL cell was designed and a complete layout (including the programming devices) was created using the TSMC 40nm LP library.  The flash transistor models were obtained from \cite{Abusultan_ICCD_2016} and were suitably modified to reflect the characteristics and variations of the TSMC 40nm library. The design rules for the flash transistors were obtained from ITRS. The layout of the FTL cell was created as a standard cell with an area of 15.6 $\mu m^2$. For reference, if X represents the drive strength, an X4 DFF and an X4 NAND gate have an area of $5.6 \mu m^2$ and 2.8 $\mu m^2$ respectively, while their delay optimized X8 counterparts have an area of 14.347 $\mu m^2$ and 7.3 $\mu m^2$ respectively. The \{setup, C2Q\} of a X4 DFF is \{67ps, 168ps\}.


	
	There are a total of 117 distinct threshold functions of 5 or fewer variables.  A numbered list of these is given in~\cite{book:muroga} and can also be accessed at~\cite{Threshold5List}.  In this section, we use the same numbering scheme as in~\cite{book:muroga} to identify the functions. In the sequel, the FTL cell trained to implement the threshold function numbered $n$ in~\cite{book:muroga} will be referred to as $FTL_{n}$, and the corresponding CMOS implementation will be denoted as $CMOS_{n}$.  The threshold function itself will be denoted as $F_{n}$. 
	
	\subsection{\textbf{Training Iterations}}
	
\begin{figure}[h]
		\vspace*{10pt}
		\centering
		\includegraphics[width=0.9\columnwidth]{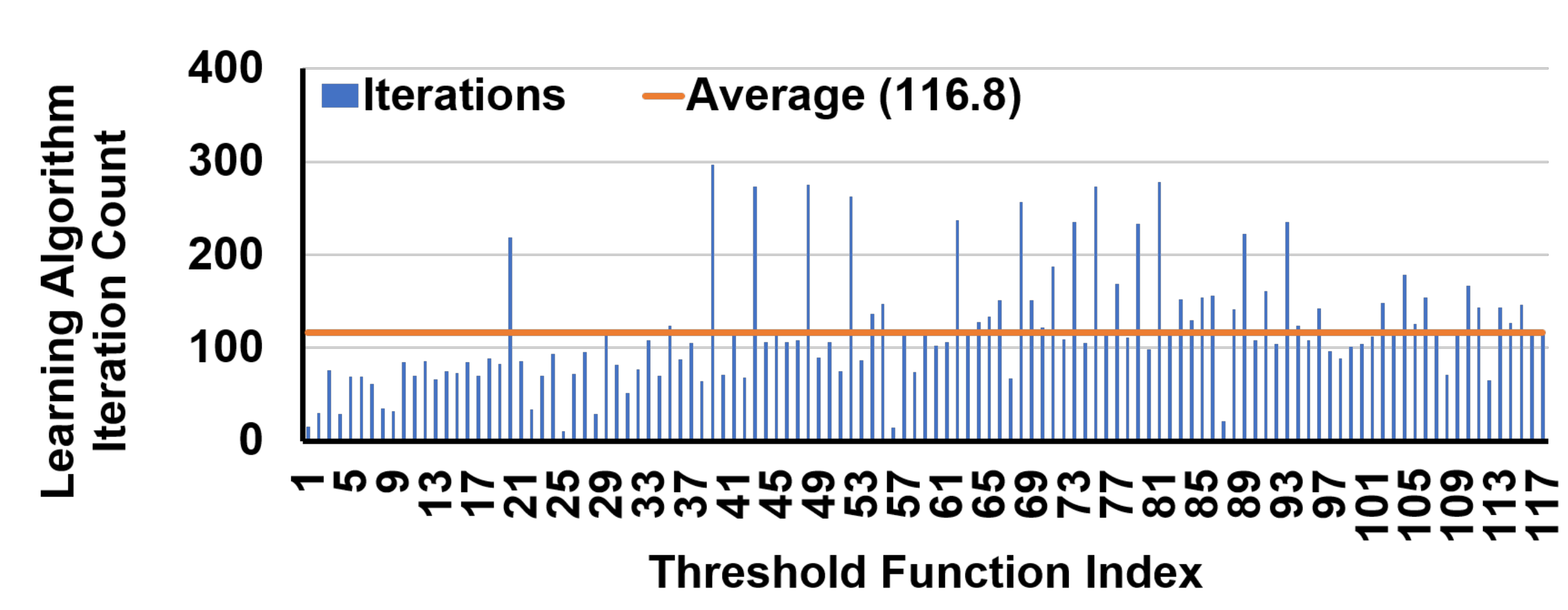}
		\caption{\small Iteration count for the modified perceptron learning algorithm for all 117 functions of 5 or fewer variables.}
		\label{fig:iteration_count}
	\end{figure}

	The modified PLA algorithm was used to train the FTL cell for robustness (see Section~\ref{robustness_training}) for all 117 functions.  Figure~\ref{fig:iteration_count} shows the number of iterations needed for training for each of the 117 functions. The actual number of iterations were about 10X lower than the theoretical upper bound, presented in Section \ref{sec:PLA}.
	
	\subsection{\textbf{Area, Delay and Power Comparison}}
	
\begin{figure}[h]
		\vspace*{10pt}
		\centering 
		\includegraphics[width=0.9\columnwidth]{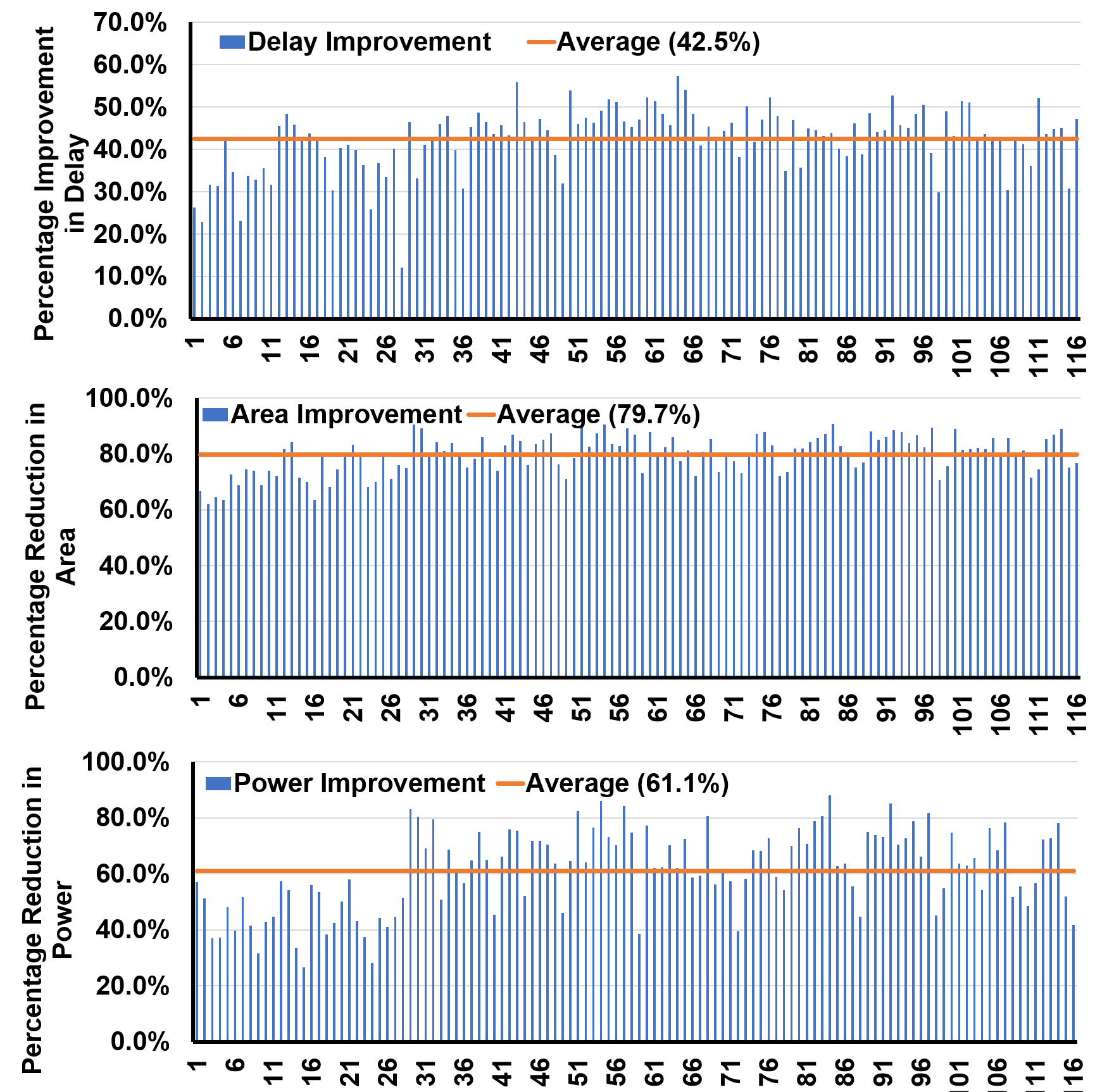}
		\caption{{\small PPA improvements of FTL over CMOS implementations.}}
		\label{fig:combined_improvement} 
		\vspace{10pt}
	\end{figure}

	Each of the 117 functions were implemented as FTL cells, and also synthesized by Cadence Genus\textsuperscript{\textcopyright}
	and placed and routed using Cadence Innovus\textsuperscript{\textcopyright}
	, using the TSMC 40nm LP standard cells. The total delay (logic delay + setup time + clock-to-Q delay) and power values were determined by simulating the circuits at 25$^{\circ}$C at 20\% input switching activity.  Figure~\ref{fig:combined_improvement} shows that each of the FTL implementations of the 117 functions have substantially  smaller area, power and delay when compared to the CMOS equivalent. The averaged improvements of FTL over CMOS are: \textbf{area (79.5\%)}, \textbf{delay (42.5\%)} and \textbf{power (61.1\%)}. 
	
\begin{figure}[h]
		\vspace{7pt}
		\centering
		\includegraphics[width=0.9\columnwidth]{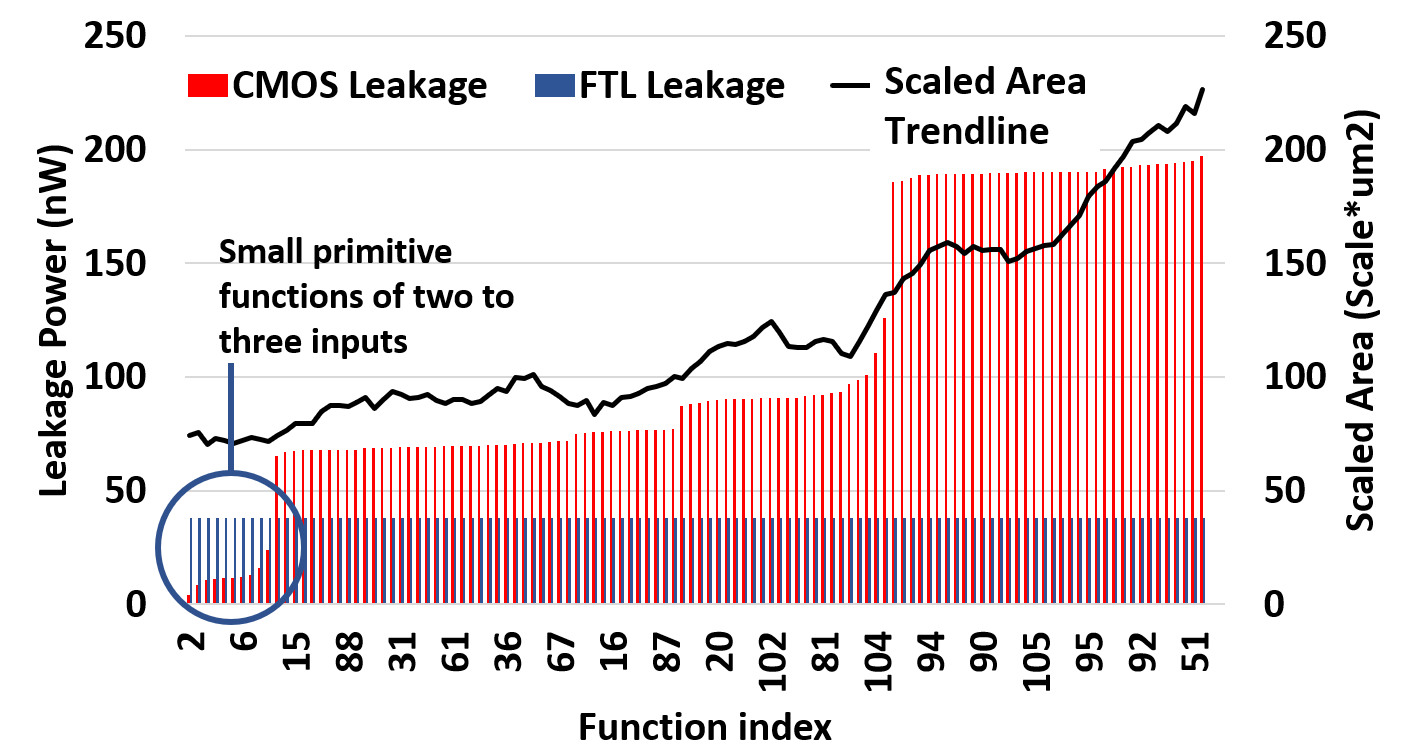}
		\vspace{-5pt}
		\caption{\small Leakage power of FTL versus CMOS implementations.}
		\label{leakage}
	\end{figure}

	Figure \ref{leakage} compares the leakage power of the FTL and CMOS implementations of the 117 functions.  The functions are arranged in ascending order of their CMOS leakage values.  Unlike the CMOS implementations, the leakage power of the FTL implementations is nearly constant. Also plotted is the area trend line of CMOS implementations, to illustrate the strong correlation of leakage power with area. The few FTL implementations that had higher leakage (shown circled) were all small logic primitives. Nevertheless, the \textit{total} power (see Figure~\ref{fig:combined_improvement}) of  FTL implementations of even these functions is far less than the CMOS implementations. These functions can be avoided if leakage minimization is the primary design goal. 
	
	\subsection{\textbf{Experiments on Training for Robustness}}
	\label{subsec:expResultsRobustness}
	
	This experiment demonstrates the robust PPA training method, described in Section~\ref{robustness_training}, to improve yield.  The test function chosen was $F_{115} = [\bm{W}; T] = [4, 1, 1, 1, 1; 5] =ab+ac+ad+ae$. The experiment consisted of training multiple versions of $FTL_{115}$ for various values of the parasitic capacitances $C_1$ and $C_0$, and for each solution, performing 100K Monte Carlo simulations with local and global process variations\footnote{Several dozen parameters are varied in the HSPICE models provided by the vendor}, and checking if the truth table was correctly realized.  
	Table~\ref{fig:rob_training_cap} shows the delay and yield for various values of $C_1$ and $C_0$. The functional yield was improved from 13\% to 100\% (i.e. truth tables of all 100K instances were verified to be correct) by increasing the values of $C_1$ and $C_0$. There are two important observations to be made here. First, even though the weights of $b$, $c$, $d$, $e$ are equal, the corresponding flash transistors received different threshold voltages ($V_2, V_3, V_4, V_5$).  This shows that the perceptron learning algorithm, working in concert with HSPICE, accounts for the layout parasitics. Second, the delay \textit{improves} with increasing robustness, due to the increase in the difference between the voltages at nodes N5 and N6,(see Section \ref{sec:PLA}).
	
	\begin{table}[h]
		\vspace*{15pt}
		\centering
		\caption{\small Multi-Corner Monte Carlo results with 100K simulations of $FTL_{115}$, trained for robustness using various capacitor values (fF)}
		\label{fig:rob_training_cap}
		\small
		\begin{tabular}{|c|c|c|c|}
			\hline
			$C_1,$   & Average Vt Values (V) &Yield& Delay\\
			$C_0$ & ($V_1, V_2, V_3, V_4, V_5; V_{l0}, V_{r0}$) & \% & (ps) \\  \hline
			0.00    & 0.64, 0.74, 0.72, 0.74, 0.72; 1.00, 0.74       &  13  &  244  \\ \hline
			0.01    & 0.62, 0.72, 0.7, 0.74, 0.74; 1.00, 0.70       &  20  & 220    \\ \hline
			0.02    & 0.58, 0.74, 0.72, 0.74, 0.72; 1.00, 0.64       &  43  & 204   \\ \hline
			0.05     & 0.48, 0.68, 0.66, 0.70, 0.66; 1.00, 0.56     &  59 & 162    \\ \hline
			0.10    & 0.34, 0.56, 0.54, 0.60, 0.62; 1.00, 0.46      & 100  & 138    \\ \hline    \end{tabular}
		\vspace{10pt}
	\end{table}
	
	\begin{figure}[h]
		\vspace*{5pt}
		\centerline{\includegraphics[width=0.85\columnwidth]{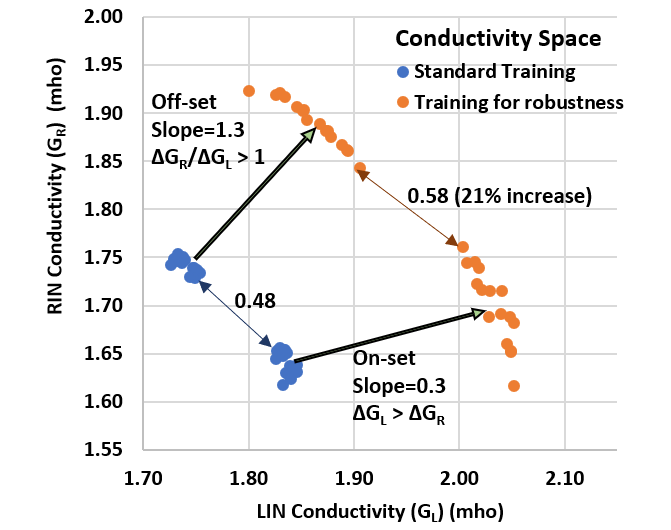}}
		\caption{\small Conductivity $G_L$ and $G_R$ of  $FTL_{115}$ [TT, 0.9V, 25$^{\circ}$C]. In the conductivity space, gap between off-set minterms and on-set minterms increases, when the training is done for robustness.}
		\vspace{10pt}
		\label{fig:conductivity_space_real}
	\end{figure}
	
	In Section~\ref{robustness_training} we argued that training an FTL cell with a \textit{handicap} in the form a parasitic capacitance on N1 and N2 will improve the robustness by increasing the smallest gap in conductance between the LIN and the RIN.  Figure~\ref{fig:conductivity_space_real} demonstrates this very important characteristic of the robust PPA algorithm for FTL. It is a plot of the \textit{conductivity space}, i.e., $G_R$ versus $G_L$, of an FTL when trained for the test function $F_{115}$, with and without the parasitic (handicap) capacitances.  The blue points (orange points) correspond to the $G_L$ and $G_R$ values of the on-set and off-set minterms of $F_{115}$ in the absence (presence) of the parasitic capacitances $C_{1}$ and $C_{0}$ ($C_1$ = $C_0$ = 0.1fF).

	Recall that for an on-set minterm $G_L > G_R$ and for an off-set minterm, $G_R > G_L$. The plot clearly demonstrates that training with the parasitic capacitances dramatically improves the robustness in two ways.  First, there is a significant increase (by 21\%) in the shortest distance between the two closest on-set and off-set minterms, as indicated in Figure \ref{fig:conductivity_space_real}. Second, the increase in $G_L$ is greater than the increase in $G_R$, i.e. $\Delta G_L/ \Delta G_R > 1$ for the on-set minterms, and vice-versa for the off-set minterms.  Both of these effects contribute to reducing the contention in the sense amplifier in deciding the function output, which in turn directly improves the speed as well, resulting in higher robustness \underline{and} higher performance. 
	
	
	\subsection{\textbf{Delay Distributions}}
	\label{subsec:DelayDistributions}
	
\begin{figure}[h]
		\vspace*{10pt}
		\centering
		\includegraphics[width=0.9\columnwidth]{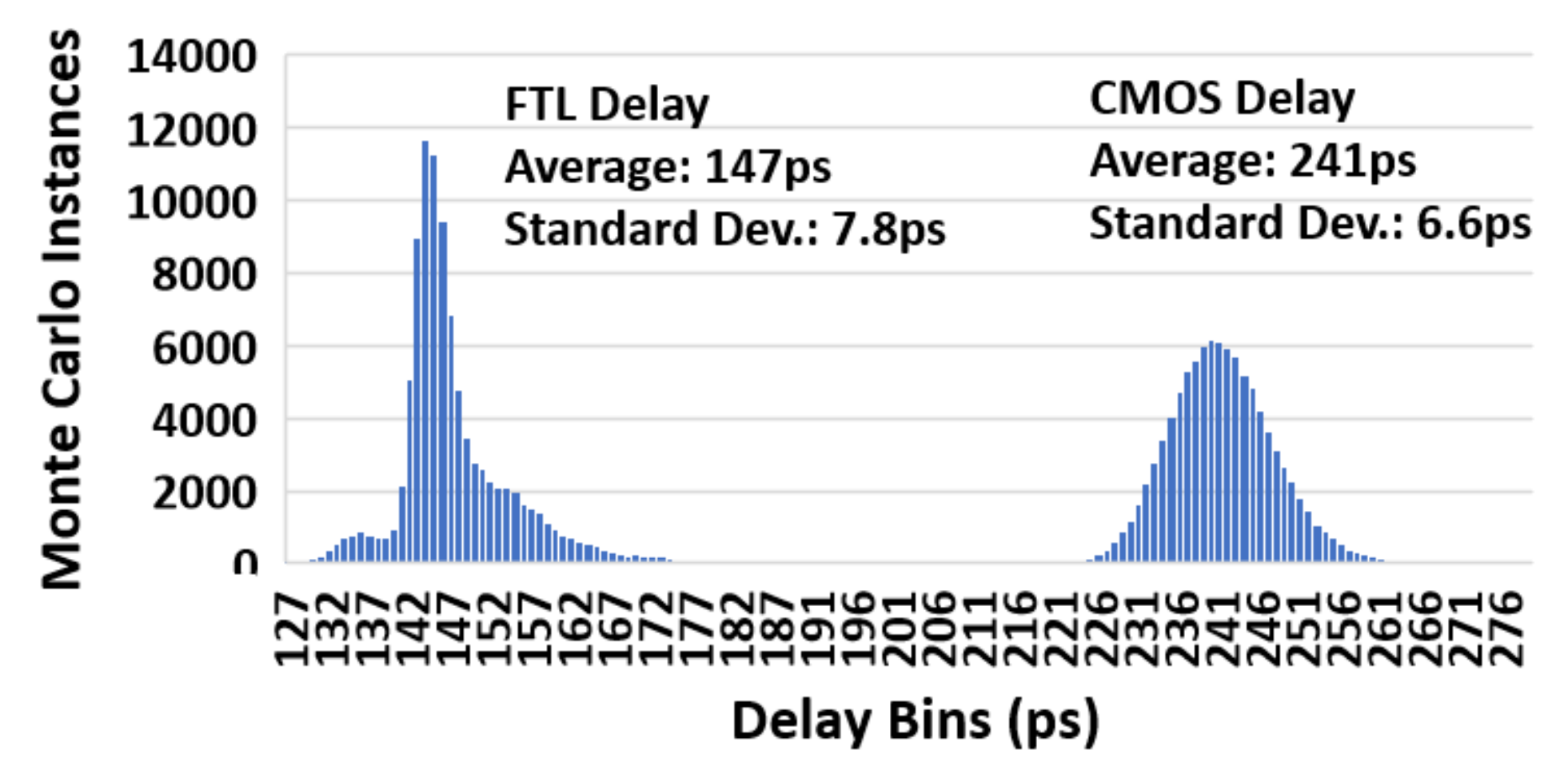}
		\caption{\small Delay histogram of $FTL_{115}$ and $CMOS_{115}$ with 100K Monte Carlo simulations. $PVT = [TT,0.9 V,25^\circ C]$.}
		\label{monte_carlo_delay}
	\end{figure}

	This experiment compares the distributions of delays of FTL and CMOS implementations. We show the results for the function $F_{115} = [\bm{W}; T] = [4, 1, 1, 1, 1; 5]$.  The PVT corner setting was $[P,V,T] = [TT,0.9V,25^\circ C]$.  100K Monte Carlo instances were generated for both $FTL_{115}$ and $CMOS_{115}$.  The function of each of the 100K FTL instances was verified against the truth table for correctness, for both $FTL_{115}$ and $CMOS_{115}$.  The histograms of delays are shown in Figure~\ref{monte_carlo_delay}. These clearly demonstrate the delay advantage of the FTL cell over its CMOS equivalent, even in the presence of process variations.  The difference in standard deviation between the two is insignificant. Note that the FTL instances with large delays can be \textit{re-programmed} to further reduce the delay.  This capability is not possible for the CMOS versions. 
	
	\begin{table}[h]
		\vspace*{20pt}
		\centering
		\caption{\small Delay, total power and power-delay-product (PDP) of $FTL_{115}$, trained at $V_{DD} = 0.9V$, and $C_{0} = C_1 = 0.1fF$.}
		\label{voltage_sweep2}
		\small
		\begin{tabular}{|c|c|c|c|c|}
			\hline
			\begin{tabular}[c]{@{}c@{}}Supply\\ Voltage (V)\end{tabular} & \begin{tabular}[c]{@{}c@{}}Flash Gate\\ Voltage (V)\end{tabular} & Power (u) & Delay (ps) & PDP    \\ \hline
			0.8         & 0.8                                                              & 14.3      & 198.1      & 2837.1 \\ \hline
			0.85        & 0.825                                                            & 20.5      & 157.6      & 3228.7 \\ \hline
			0.9         & 0.85                                                             & 26.1      & 130.2      & 3396.9 \\ \hline
			0.95        & 0.875                                                            & 40.3      & 111.2      & 4482.7 \\ \hline
			1           & 0.9                                                              & 53.1      & 97.0       & 5148.6 \\ \hline
			1.05        & 0.925                                                            & 76.0      & 86.4       & 6562.9 \\ \hline
			1.1         & 0.95                                                             & 85.0      & 78.2       & 6644.0 \\ \hline
		\end{tabular}
		\vspace{5pt}
	\end{table}
	\subsection{\textbf{Dynamic Voltage Scaling}}
	\label{subsec:VoltageScaling}
	
	Voltage scaling is a common mechanism to trade off performance against power. Table~\ref{voltage_sweep2} shows the results of training $FTL_{115}$ at $0.9V$.  The FTL was programmed with the resulting set of flash threshold voltages, and then operated over the voltage range $[0.8V, 1.1V]$. To ensure proper operation across all voltages, the gate voltages of the flash transistors were also scaled in this experiment. This result demonstrates how a single $\bm{V}_T$ assignment can be used for dynamic voltage scaling. Note that the delay varies by 2.5X, power varies by 5.9X and the PDP (energy) varies by 2.3X, as the supply voltage varies over [0.8V, 1.1V]. This shows that the FTL cells offer a healthy power, delay, and energy tradeoff by voltage scaling.

	\subsection{\textbf{Post-fabrication Timing Correction}}
	The experiments described in Sections \ref{subsec:expResultsRobustness}, \ref{subsec:DelayDistributions} and \ref{subsec:VoltageScaling} all point to the flexibility of FTL due to its unique characteristic of allowing for programming of the flash transistor threshold voltages after fabrication. It should come as no surprise then that this can also be used to correct timing errors. 
	
	\begin{figure}[h]
		\centering
		\includegraphics[scale=0.45]{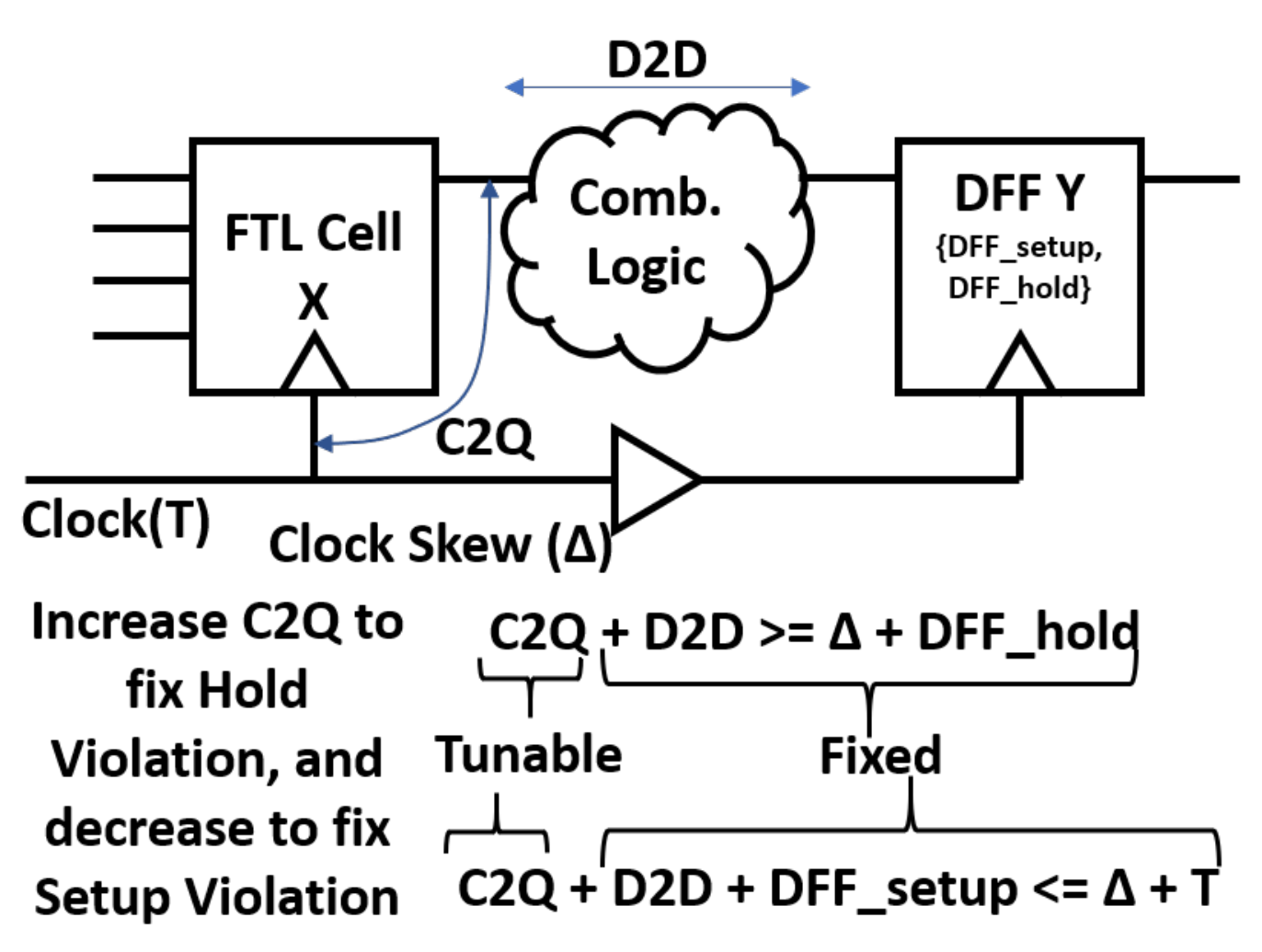}
		\caption{\small Datapath to demonstrate post-fab timing corrections}
		\label{fig:setup_hold_test_circuit}
	\end{figure}
	
	\begin{figure}[h]
		\vspace*{20pt}
		\centering
		\includegraphics[width=\columnwidth]{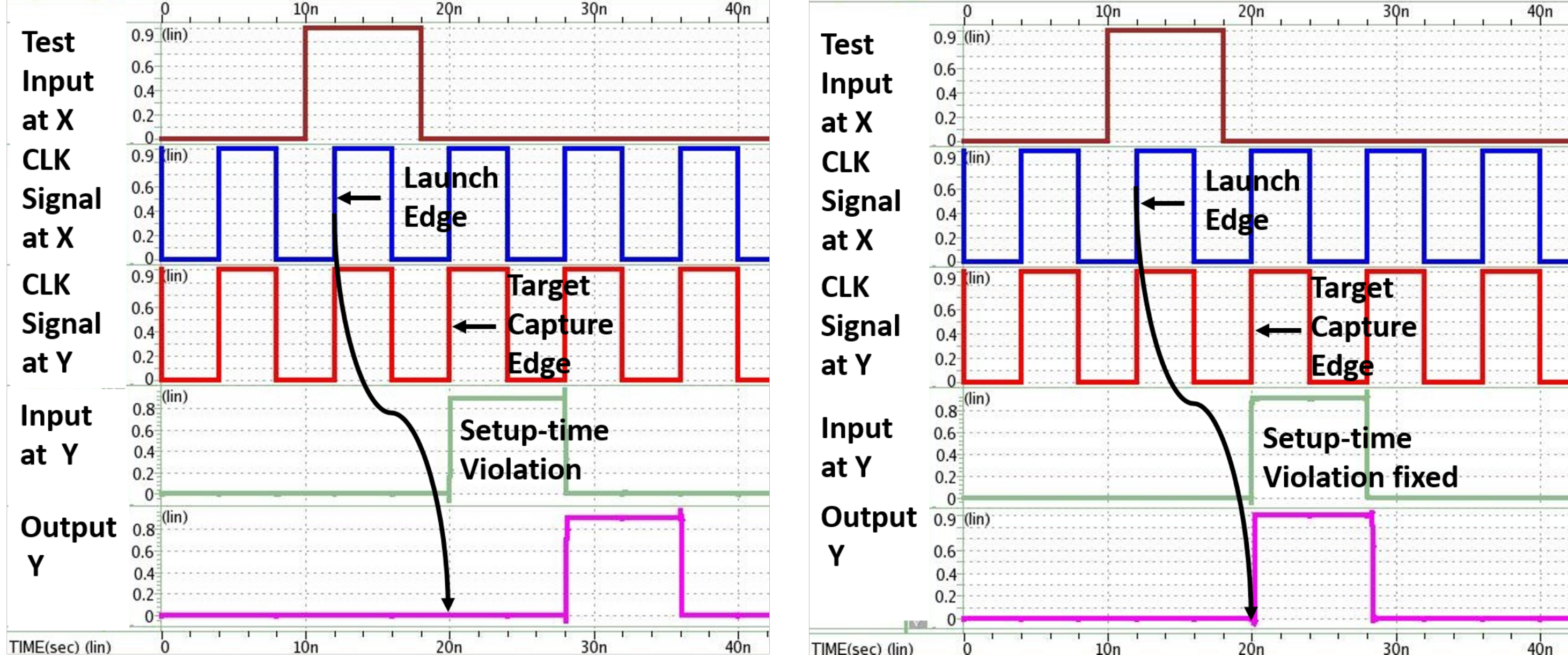}
		\caption{\small Correcting setup time violation with an FTL cell after fabrication. C2Q of FTL cell reduced from 180 ps to 142 ps.}
		\label{setup_fix_waveform}
		\vspace*{10pt}
	\end{figure}
	
	\begin{figure}[h]
		\vspace*{15pt}
		\centering
		\includegraphics[width=\columnwidth]{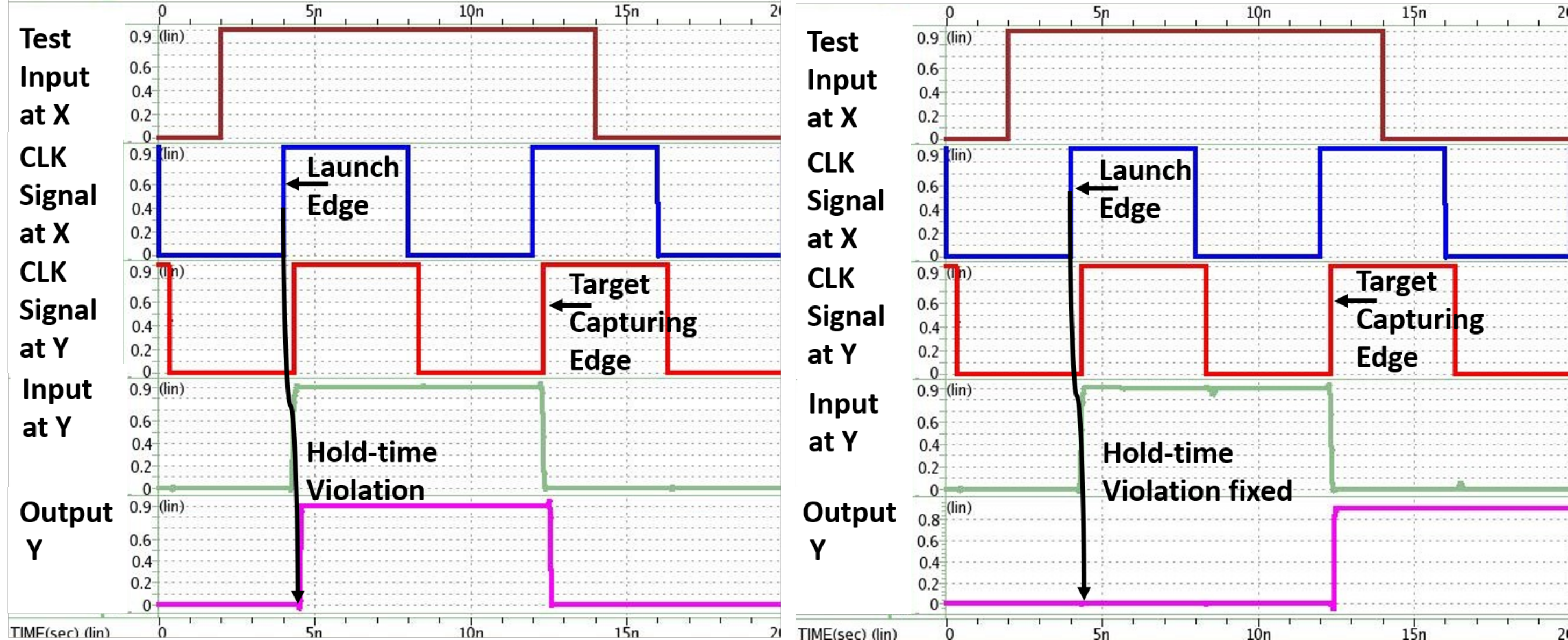}
		\caption{\small Correcting hold time violation with an FTL cell after fabrication. C2Q of FTL cell increased from 142 ps to 180 ps.}
		\label{hold_fix_waveform}
	\end{figure}
	~
	
	Figure~\ref{fig:setup_hold_test_circuit} shows a small datapath that was constructed to demonstrate how setup time and hold time violations can be corrected after fabrication in an FTL design. The datapath consists of clock-to-Q (C2Q) delay, combinational delay (D2D) and DFF specifications for setup ($DFF\_setup$) and hold ($DFF\_hold$) times. The clock is skewed by an appropriate amount $\Delta$, to generate either a setup time or a hold time violation. The violations are corrected by reprogramming the FTL cell to produce different C2Q values. 
	
	Figure \ref{setup_fix_waveform} shows how the data launched from FTL X misses the target clock edge at DFF Y, thereby violating setup time. To fix the setup time, the C2Q of FTL X is decreased. Similarly, Figure \ref{hold_fix_waveform} shows how the data launched at FTL X gets captured by DFF Y one cycle early, thereby overwriting the old value at DFF Y. By increasing the C2Q value of the FTL X, the old value at the input of Y is retained for a longer time, which satisfies the hold time condition. Since the FTL cells are programmed post-fabrication, the delay can also be modified after fabrication. Using the same idea of post-fabrication $V_T$ adjustment, an FTL cell can be reprogrammed to mitigate delay increases due to aging.

	\subsection{\textbf{Chip-level programming architecture}}
	
	Although the full architecture for programming the FTL cells in an ASIC is not presented here, this section describes the programming architecture in brief. An on-chip decoder architecture is used to address the flash transistors of the FTL cells, during programming. The address for the decoder is sent into the chip using a serial communication protocol along with a programming clock.  The high voltage line needed for sending programming pulses to flash transistors is generated and sent into the chip using an off-chip voltage source. The pin count overhead for programming is low (only 3 pins are needed). When the address is received, the decoder activates a specific flash transistor of a specific FTL cell for programming.
	
	\nocite{Neutzling_ICCAD_2015,Tragoudas:1999,dft,wseas,Neutzling_TCAD_2018}
	
	
	\section{\textbf{Related work}}
	\label{sec:RelatedWork}
	
	\subsection{\textbf{Threshold Logic}}
	
	The study of threshold functions and the development of threshold gates date back to the 1960s culminating in the authoritative book by Muroga~\cite{book:muroga}.  Since then, an extensive body of theoretical work, new circuit architectures and implementations have been published.  References~\cite{Beiu_IJCNN_2003} and ~\cite{Celinski_2003} provide a detailed survey of work prior to 2003. 
	
	One of the earliest reported works that demonstrated the operation of threshold logic gates using \textit{flash transistors} was reported in ~\cite{Bohossian_ISIS_1997, Rodriguez-Villegas_ISCAS_2002}.    It was an analog design of a single cell to demonstrate proof of concept.  The focus has shifted to exploring the use of emerging devices such as RRAMs, STT-MTJs, and others,  to implement threshold gates ~\cite{Savas_SSE_2007, Yang_NANOARCH_2014, Mozaffari_TNANO_2018}. Several recent works have devised efficient algorithms for determining weights aimed at robust threshold gates~\cite{Mozaffari_TCSI_2018, Mozaffari_TNANO_2018}.  However until recently, due to the lack of designs tools and incompatibility with existing design methodologies, threshold logic remained outside mainstream VLSI design.  
	
	Recently, \cite{Kulkarni_TVLSI_2016} reported an architecture of a threshold gate and showed how it can be integrated with the standard-cell ASIC design methodology using commercial tools.  In addition, they reported significant improvements in PPA of an actual silicon implementation of ASIC with threshold gates~\cite{Jinghua:CICC2015}.  Their architecture, however, severely limits the number of threshold functions that can be implemented.   This is because the weight $w_i$ associated with input $x_i$ is implemented by using $w_i$ transistors each driven by signal $x_i$.  Hence, their circuit has severe fan in limitations.  For instance, the design in~\cite{Kulkarni_TVLSI_2016} can only realize 11 of the 5-input threshold functions, whereas, as demonstrated here, the FTL-5 cell can realize all 117 functions  In addition, representing weights using multiple transistors significantly reduces the robustness and prevents it from scaling to lower geometries.  Finally, the FTL cell is programmed after fabrication, preventing copying by a foundry, and numerous opportunities to correct failures and tune for high performance and aging effects.  
	
	\subsection{\textbf{Flash Technology}}
	
	Many research efforts have studied flash devices and their use in memory. A short list includes~\cite{An_ICNIDC_2010, Choi_IEDM_2012}.  These papers report details of flash devices and their characterization. However, they do not describe the use of flash transistors for logic circuits. A good deal of work in flash has been reported in the area of architectural techniques to increase flash memory endurance. Some representative works include wear leveling techniques, which are used in flash-based memory blocks~\cite{Qureshi_MICRO_2009}, to compensate for the fact that flash transistors typically have a finite (10k - 100k) number of times they can be written~\cite{Jung_CASES_2007, Boboila_2010}.  In traditional flash memory, \textit{wear leveling} is performed at the architectural level to spread the wear of the cells. 
	
	The authors of~\cite{Abusultan_ICCAD_2016} present a design flow to implement flash-based digital circuits at the block level.  These efforts present results for a programmable logic array style cell design and illustrate its use in a modified standard-cell style VLSI design flow. In contrast, the work of this paper focuses on threshold logic and is envisioned for use in a traditional standard-cell based flow. An FTL cell can replace a D~flip-flop and some or part of its logic cone in any CMOS netlist.  
	
	To the best of our knowledge, there has been no work prior to this paper which describes the synthesis, detailed electrical characterization of sequential flash-based threshold logic cells.

	\section{Conclusion}
	\label{sec:Conclusions}
	
	In this paper, we proposed a novel threshold logic cell (FTL) using flash transistors. A modified perceptron learning algorithm was also proposed to program the FTL cell. Substantial area (79.7\%), power (61.1\%) and performance 42.5\%) improvement of the FTL cells was demonstrated against their conventional 40nm standard-cell based designs of the same functions. By adding a capacitor to introduce a handicap in the FTL cell during simulation, this paper shows that the learning algorithm counters the effect of the handicap by generating more robust solutions. Robustness against PVT variations was demonstrated using 100K Monte Carlo simulations, demonstrating a 100\% yield. We also demonstrated that FTL cells are amenable to dynamic voltage scaling, and post-silicon tuning of setup and hold time violations.

	%


	\nocite{Mozaffari_TNANO_2018,Rajendran_NANOARCH_2010, ROSE_IEEE_2012, Soltiz_IEEE_TRANS_2013, Soltiz_NANOARCH_2012, Detorakis_NeuralAS_2017, Uddin_2018, Sayyaparaju_GLVLSI_2017,Yao_Nanotech_2012,Patil_ICCD_2010}
	\newpage
	\bibliographystyle{IEEEtran}
	\bibliography{IEEEabrv,ICCD2019-main-4-aw_arxiv}
\end{document}